\newif\ifjsait

\jsaittrue
\newif\ifdouble

\doubletrue
\newif\ifshort
\shortfalse

\newif\ifveryshort
\veryshortfalse

\ifjsait
\documentclass[journal,12pt,onecolumn]{IEEEtran}
\usepackage[letterpaper, top=1in, bottom=1in, left=1in, right=1in]{geometry}
\usepackage{setspace}

\else
\documentclass[conference,a4paper]{IEEEtran}
\fi
\usepackage{footnote}
\usepackage{footmisc} 
\usepackage{amsthm}
\usepackage{latexsym}
\usepackage{amssymb}
\usepackage{graphicx}
\usepackage{caption}
\usepackage{subcaption}
\usepackage{amsmath}
\usepackage{color}
\usepackage{cite}
\usepackage{cleveref}
\usepackage[english]{babel}
\usepackage[utf8]{inputenc}
\usepackage[T1]{fontenc}
\usepackage{url}
\usepackage{ifthen}
\usepackage{tabularx}

\theoremstyle{plain}
\newtheorem{theorem}{Theorem}
\newtheorem{lemma}{Lemma}

\newtheorem{corollary}{Corollary}
\theoremstyle{definition}
\newtheorem{definition}{Definition}
\theoremstyle{remark}

\newcommand{\off}[1]{}

\ifCLASSINFOpdf
\else
\fi
\begin{document}
\title{Secure Adaptive Group Testing\thanks{Parts of this work were presented at the IEEE International Symposium on Information Theory, ISIT 2018.}}
 \author{%
   \IEEEauthorblockN{Alejandro Cohen \IEEEauthorrefmark{1} \hspace{15 mm}
                     Asaf Cohen \IEEEauthorrefmark{2} \hspace{15 mm}
                     Omer Gurewitz\IEEEauthorrefmark{2}}
                     
   \IEEEauthorblockA{\IEEEauthorrefmark{1}%
                     RLE, MIT, Cambridge, MA, USA, cohenale@mit.edu} 
                     
   \IEEEauthorblockA{\IEEEauthorrefmark{2}%
                     School of Electrical and Computer Engineering, Ben-Gurion University, Israel,
                    \{coasaf, gurewitz\}@bgu.ac.il}
 }

\maketitle
\begin{abstract}
\emph{Group Testing} (GT) addresses the problem of identifying a small subset of defective items from a large population, by grouping items into as few test pools as possible. In \emph{Adaptive GT} (AGT), outcomes of previous tests can influence the makeup of future tests. Using an information theoretic point of view, Aldridge $2012$ showed that in the regime of a few defectives, adaptivity does not help much, as the number of tests required is essentially the same as for non-adaptive GT.

\emph{Secure GT} considers a scenario where there is an eavesdropper who may observe a fraction $\delta$ of the tests results, yet should not be able to infer the status of the items. In the non-adaptive scenario, the number of tests required is $1/(1-\delta)$ times the number of tests without the secrecy constraint.

In this paper, we consider \emph{Secure Adaptive GT}. Specifically, when during the makeup of the pools one has access to a private feedback link from the lab, of rate $R_f$. We prove that the number of tests required for both correct reconstruction at the legitimate lab, with high probability, and negligible mutual information at the eavesdropper is $1/min\{1,1-\delta+R_f\}$ times the number of tests required with no secrecy constraint. Thus, unlike non-secure GT, where an adaptive algorithm has only a mild impact, under a security constraint it can significantly boost performance. A key insight is that not only the adaptive link should disregard the actual test results and simply send keys, these keys should be enhanced through a ``secret sharing" scheme before usage. We drive sufficiency and necessity bounds that completely characterizes the Secure Adaptive GT capacity. 

Moreover, we consider additional models of Secure Adaptive GT, where we make a clear distinction between the lab performing the tests, and the doctor analyzing the results. Specifically, we consider curious but non-malicious, non-cooperating labs. Each lab gets a fraction $\delta$ of pool-tests to perform. Yet, we want to keep each lab ignorant regarding the status of the items. In contrast, the doctor who gets all outcomes, should successfully decode. When there is a feedback from each lab, we show that even if a curious lab obviously sees its own feedback (i.e., it is locally-public to Eve), secure adaptive GT is still possible, and at a rate that can be equal to the one without a security constraint at all, by an application of the Leftover Hash Lemma, using the data of one lab to protect against another. 
\end{abstract}

                                   %
\section{Introduction}
\emph{Group Testing} (GT) was introduced in a seminal study by Dorfman to identify syphilis infected draftees while dramatically reducing the number of required assays~\cite{dorfman1943detection}. Specifically, the objective of GT is to identify a small subset of $K$ unknown defective items within a much larger set of $N$ items, conducting as few measurements $T$ as possible.

This problem has been analyzed in various scenarios \cite{du2000combinatorial}, one of which, \emph{Non-Adaptive Group Testing} (NGT), is when the entire pooling strategy is decided on beforehand. This scenario has also been formulated as a channel coding problem, e.g.,  \ifveryshort\cite{atia2012boolean}\else\cite{atia2012boolean, malyutov2013search, aydinian2013information}\fi, where each item is associated with a codeword, defining the pool-tests in which it participates. Recently GT was considered as well for testing coronavirus disease 2019 (COVID-19) using significantly fewer tests that the number of the tested items \cite{narayanan2020accelerated,theagarajan2020group,seong2020group,shental2020efficient,petersen2020practical,ben2020large}. In particular, for COVID-19, due to practical constraints on the maximum number of pool-test which can be analyzed together, GT is used in many cases in batches of $T$ tests each \cite{shental2020efficient,petersen2020practical,ben2020large}, when $T$ is not necessarily very large. For example, in \cite{shental2020efficient}, the number of pool-tests is 48.

\emph{Secure GT} protects the items’ privacy such that an eavesdropper who may observe only a fraction of the pool-tests, will not be able to infer the status of any of the items (negligible mutual information between the captured pool-tests and the status of the items). \cite{cohen2016secure} addressed the Secure Non-adaptive Group Testing (SNGT) model. In order to confuse the eavesdropper, instead of each item having a fixed test vector, determining in which pool-tests it should participate, each item has a vector chosen at random from a known set of vectors. The random selection was known only to the mixer. \cite{cohen2016secure} proved that when the fraction of tests observed by the eavesdropper (Eve) is $0 \leq \delta <1$, the number of tests required for both correct reconstruction at the legitimate lab and negligible mutual information at Eve's side is $1/(1-\delta)$ times the number of tests required with no secrecy constraint. Thus, the solution in \cite{cohen2016secure} relied on \emph{information theoretic security}, which offers security at the price of rate, or, in the GT case, \emph{trades security and the number of tests}.
\begin{figure}
  \centering
  \ifjsait
  \includegraphics[trim=0cm 0cm 0cm 0cm,clip,scale=1.2]{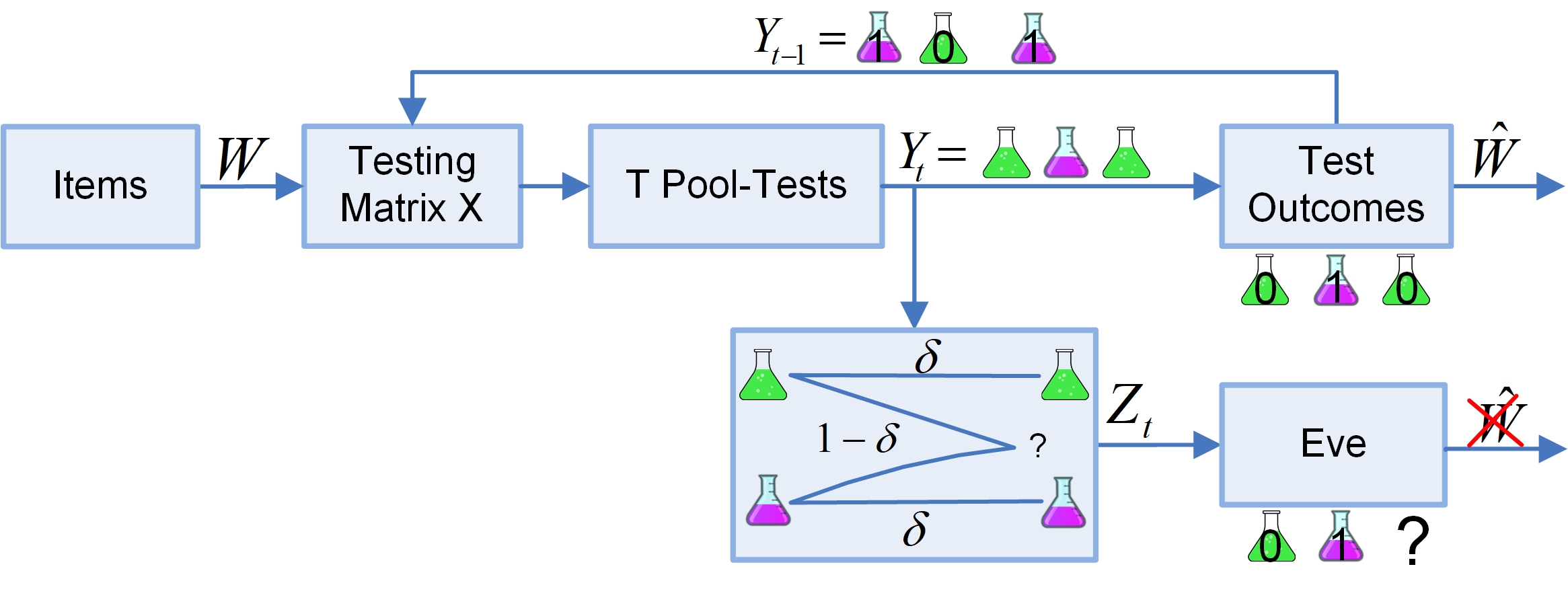}
  \else
  \includegraphics[trim=0cm 0cm 0cm 0cm,clip,scale=0.83]{fig/SecureGroupTestModelE.jpg}
  \fi
  \caption{An analogy between a channel with feedback and Secure Adaptive Group Testing. Defective items can be considered as the "transmitting" users. The others do not affect the output. The index of the defective items is denoted by $W$. The content of the testing matrix defines the codewords to be transmitted. Due to the adaptive link, this content may depend on the output of the channel - previous tests results.}
  \label{figure:group_testing_model}
\end{figure}

In this paper, we consider \emph{Adaptive Group Testing} (AGT), in which the outcomes of previous tests can influence the construction of future pools. In general, adaptivity may facilitate both a reduced number of tests ($T$) and efficient decoding techniques \cite{hughes1994two}. However, it has been shown that in many interesting cases, the decrease in $T$ is negligible. Several studies have analyzed AGT as a channel coding problem with feedback (e.g., \off{\ifveryshort\cite{aldridge2012adaptive, kealy2014capacity, aeron2010information}\else}\cite{aldridge2012adaptive, aksoylar2014information, baldassini2013capacity, kealy2014capacity, aeron2010information}\off{\fi}). Utilizing the same techniques as Shannon’s seminal study \cite{shannon1956zero} which proved that feedback does not increase channel capacity, the authors in \cite{aldridge2012adaptive} showed that feedback from the lab does not decrease the number of tests required significantly. That is, if $T$ is the minimal number of tests required for reconstruction with negligible error, when $N\rightarrow \infty$ the gain due to the adaptivity is marginal.\off{$\lim_{T\rightarrow \infty} \frac{1}{T}\log\binom{N}{K}$ is the same with or without adaptive testing.} Even with zero-error, the difference in $T$ is only between $O(K^2\log N/\log K)$ and $O(K\log N)$ \cite{du2000combinatorial}.

Surprisingly, we show that in secure-adaptive group testing, the adaptive feedback link may decrease the number of tests required significantly, down to coinciding with the number of pool-tests required with no secrecy constraint at all. While from an information theoretic perspective, previous results indeed show that in communication feedback can increase secrecy capacity \cite{ahlswede2006transmission, C8,cohen2016wiretap}, herein, shuch techniques do not apply directly, as the test transferred are physical entities, which cannot be, e.g., one-time-padded, and, as we explain in the sequel, the ``encoder'' in this case has no knowledge on which $K$ of the $N$ items it needs to protect. In other words, the number of "messages" that need to be protected is much larger that the true number that will be "transmitted", suggesting that any trivial solution which will equally protect all messages will fail to be tight.
\subsection*{Main Contribution}
We propose a new Secure Adaptive Group Testing (SAGT) algorithm. This algorithm significantly reduces the number of tests required, yet is sufficient for the legitimate lab to identify the defective items, and to keep an eavesdropper ignorant regarding any of the items.

The scheme is depicted in \Cref{figure:group_testing_model}. In the suggested solution, the set of indices describing the defective items takes the place of a confidential message; the testing matrix represents the design of the pools, where each row corresponds to a separate item. Each such matrix row is associated with a codeword where $1$ denotes the pool-tests in which the corresponding item participates.
The decoding algorithm is analogous to a channel decoding process, yet now the adaptive link from the lab (who examines the pools) to the mixer (who mixes the samples and creates the pooled tests) takes the place of a feedback link. The eavesdropper observation is analogous to the output of \textit{an erasure channel}, such that only part of the tests sent from the legitimate source (the mixer) to the legitimate receiver are observed by the eavesdropper. We assume a private adaptive link, which is not observable by the eavesdropper.

We use the private feedback link from the lab to the mixer in order to modify the testing matrix, in a way which can be comprehended by both the lab and the mixer, yet confuses the eavesdropper regarding which item participates in each pool-test. However, unlike wiretap channels with feedback, we cannot use the data on the feedback link directly, and \emph{must use a coding scheme}, to increase the number of keys we can generate. This is since in this case, the "encoder" is not aware of the message to be sent, hence, in a sense, has to protect all messages. We suggest a scheme such that although the keys generated are dependent, any subset which Eve eventually observes is independent and thus protected.
Accordingly, this adaptive algorithm decreases the factor $1/(1-\delta)$ given in SNGT, and one may use a smaller number of tests. In case the information rate on the feedback is equal to the rate of the eavesdropper’s observation, this factor is completely rescinded. Thus, we achieve the same sufficiency bound on $T$ as given for non-secure GT.

\begin{figure}
  \centering
  \ifjsait
  \includegraphics[trim=0cm 0cm 0cm 0cm,clip,scale=1.2]{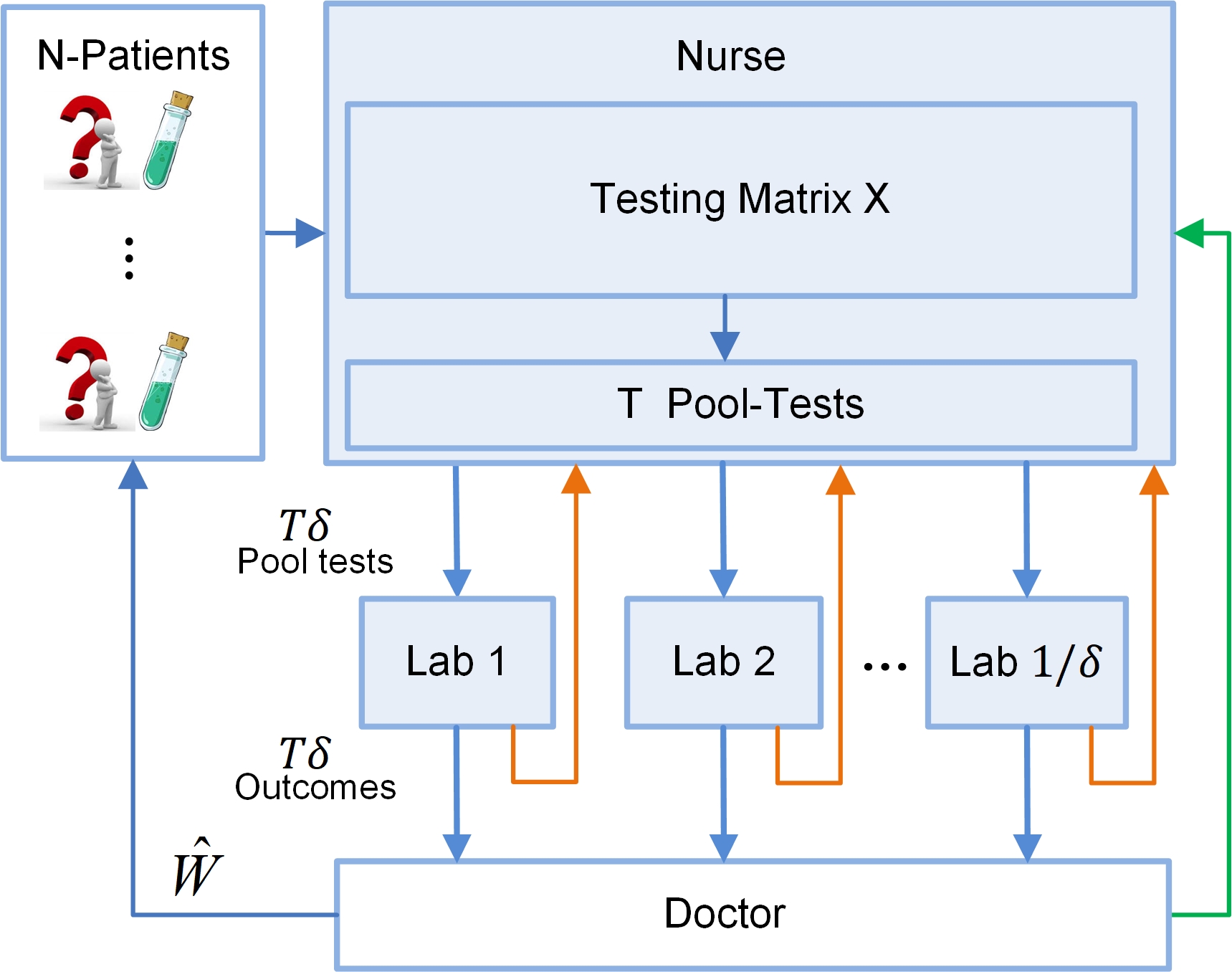}
  \else
  \includegraphics[trim=0cm 0cm 0cm 0cm,clip,scale=1]{fig/SecureGroupTestModelDoctor2.jpg}
  \fi
  \caption{Two models of secure adaptive group testing using curious but non-malicious non-cooperative labs. The green link (first model) represents the private feedback link from the doctor to the nurse while the orange links represent feedback links from the labs to the nurse, know to each lab but private from the others.}
  \label{figure:group_testing_model_doctor}
\end{figure}

In the above model, it is inconsequential if the lab performing the tests and the doctor analyzing the results are the same entity or not. In \Cref{figure:group_testing_model}, they are represented by a single block. It is important, however, that the lab is not curious (Eve is a separate entity), and the analyzer has the feedback bits when performing the analysis. In practice, it is interesting to consider two additional models, both of which including curious yet non-malicious and non-cooperative labs. Such a model is applicable if a large number of tests is collected, and to facilitate fast processing it is sent to multiple labs to work in parallel.

A graphical representation is given in \Cref{figure:group_testing_model_doctor}. Each independent lab, who tests a set of pool-tests, takes the place of the eavesdropper, in the sense that we wish to keep the specific lab ignorant regarding the items' status. A doctor receives all test results from the labs to analyze. In the first model, there is a private feedback link from the doctor to the mixer (green in \Cref{figure:group_testing_model_doctor}). In the second model, this private feedback link is not available, yet there is a direct ``locally-public" feedback\footnote{\label{public_link}Note that since the direct link is available to the lab sending it, the same lab we wish to keep ignorant, we refer to this link as ``public” in the same sense of \cite{gunduz2008secret}.} link from each lab to the mixer (orange in \Cref{figure:group_testing_model_doctor}). The content on the link is obviously known to the curious lab which sent it, hence locally-public to this eavesdropper, yet private from the others. Hence, the mixer cannot directly use the feedback data from the labs with the scheme proposed for the private link. We show that with proper precoding to extract a uniformly random and independent secret keys, using the Leftover Hash Lemma \cite{haastad1993construction,barak2011leftover}, if each lab gets a fraction $\delta \leq 1/2$ of pool-tests to analyze, the rate of the total number of tests required is \emph{the same as without any secrecy constraint}. For example, with two labs, each receiving half of the pools to test, a secure GT scheme \cite{cohen2016secure} would require twice the number of tests compared to a non-secure scheme. With our suggested locally-public feedback scheme herein, security is achieved without that multiplicative factor at all.

The rest of the paper is structured as follows. In \cref{formulation}, we formally describe the SAGT model and problem formulation. In \Cref{BooleanCompressed}, we summarize the related work. \Cref{main results} includes our main results, with the direct proved in \Cref{LowerBound} and the converse proved in \Cref{converse}. \Cref{LowerBound_moel2} describes a different secure test design, in which the makeup of the tests depends on the outcome of previous pool-tests instantaneously-one test after the other. In \Cref{sec:labs} we propose two additional models with curious but not-malicious and non-cooperative labs. \Cref{conc} concludes the paper.                                  %
\section{Problem Formulation}\label{formulation}
\begin{figure*}
  \centering
  \ifjsait
  \includegraphics[trim=0cm 0cm 0cm 0cm,clip,scale=0.86]{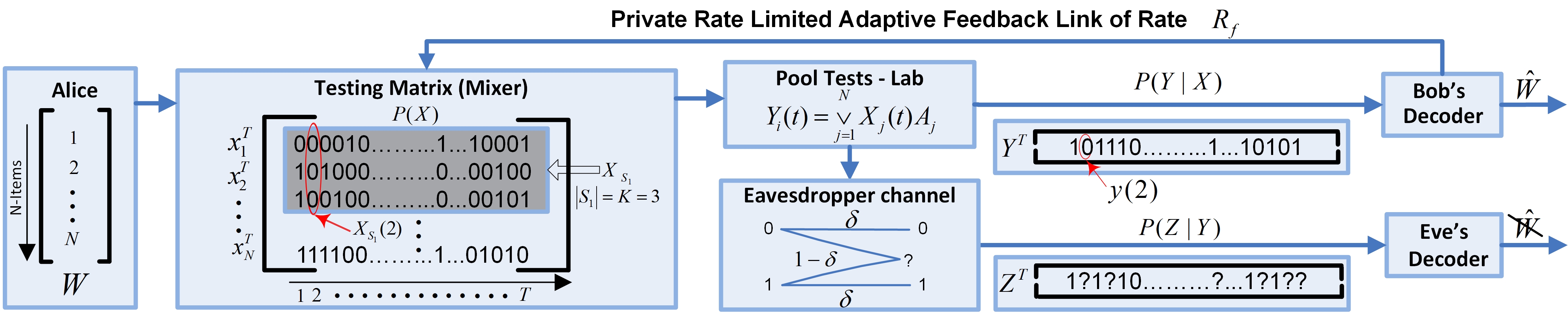}
  \else
    \includegraphics[trim=0cm 0cm 0cm 0cm,clip,scale=0.86]{fig/Ad_secure-group-testing6}
  \fi
  \caption{A secure adaptive group-testing setup.}
  \label{figure:secure-group-testing}
\end{figure*}

In \emph{SAGT}, a legitimate lab wishes to identify a small unknown subset $\mathcal{K}$ of defective items form a larger set $\mathcal{N}$, while reducing the number of measurements, $T$, as much as possible \emph{and keeping an eavesdropper}, which is able to observe a subset of the tests' results ignorant regarding the status of the $\mathcal{N}$ items. The adaptive link allows the outcomes from previous tests to influence the makeup of future tests in order to further reduce the total number of measurements. $N=|\mathcal{N}|$, $K=|\mathcal{K}|$ denote the total number of items, and the number of defective items, respectively. The status of the items, defective or not, should be kept secure from the eavesdropper, but detectable by the legitimate lab.
We assume that in each round the number $K$ of defective items is known apriori. This is a common assumption in the GT literature  \cite{macula1999probabilistic}.\footnote{However, if the number of defective items is not known apriori, \cite{damaschke2010competitive,damaschke2010bounds} provide methods/bounds on how to ``probably approximately" correctly learn the value of $K$ in a single stage with $O(\log N)$ tests.} \Cref{figure:secure-group-testing} gives a graphical representation of the model.

Throughout the paper, we use boldface to denote matrices, capital letters to denote random variables, lower case letters to denote their realizations, and calligraphic letters to denote the alphabet. Logarithms are in base $2$\off{ and $h_b(\cdot)$ denotes the binary entropy function}.

In general, GT is defined by a testing matrix
\begin{equation*}
 \textbf{X}=\{X_{j}(t)\}_{1\leq j\leq N, 1\leq t \leq T} \in \{0,1\}^{N\times T},
\end{equation*}
where each row corresponds to a separate item $j\in\{1,\ldots,N\}$, and each column corresponds to a separate pool-test $t\in\{1,\ldots,T\}$.
For the $j$-th item, $X_{j}(t)=1$ if item $j$ participates in the $t$-th pool-test and $X_{j}(t)=0$ if it does not.
Denoting by $A_j\in \{0,1\}$ an indicator function indicating whether the $j$-th item belongs to the defective set, the pool-test outcome $Y(t) \in Y^{T}$ is therefore
\ifshort
\begin{equation*}
 Y(t)=\bigvee_{j=1}^{N}X_{j}(t)A_j=\bigvee_{d\in \mathcal{K}}X_{d}(t),
\end{equation*}
\else
\begin{equation*}
Y(t)=\bigvee_{j=1}^{N}X_{j}(t)A_j=\bigvee_{d\in \mathcal{K}}X_{d}(t),
\end{equation*}
\fi
where $\bigvee$ is used to denote the boolean OR operation.

In AGT, we assume the makeup of a testing pool can depend on the outcomes of earlier tests, by a feedback link from the lab to the mixer, such that for any test $t>1$ and any item $j$, $X_{j}(t)= X_{j}(t|Y(1),\ldots,Y(t-1))$. Note, however, that the simple formulation above suggests that the mixer has \emph{all} test outcomes thus far. The model we will analyze in this paper will be, in fact, more general and the lab may send information at certain \emph{rate} $R_f$, allowing both the formulation above as well as more complicated ones.

We assume this feedback link is private.
That is, symbols $F_{t}, t \in \{1,\ldots,T\}$ are sent over a private  link, secretly from the eavesdropper.
The feedback alphabets are denoted by $\{\mathcal{F}_1,\ldots,\mathcal{F}_{T}\}$. Their cardinalities must satisfy \ifveryshort $\frac{1}{T}\sum_{t=1}^{T}\log(|\mathcal{F}_{t}|)\leq R_{f}$. \else
\ifshort
\begin{equation} \label{model 0}
 \frac{1}{T}\sum_{t=1}^{T}\log(|\mathcal{F}_{t}|)\leq R_{f}.
\end{equation}
\else
\begin{equation*} \label{model 0}
\frac{1}{T}\sum_{t=1}^{T}\log(|\mathcal{F}_{t}|)\leq R_{f}.
\end{equation*}
\fi
\fi
At each time instant $t$, $F_{t-1}$ is computed by the lab and revealed to the mixer.
The symbol $F_{t}$ at time instant $t$ may depend on $Y^{t-1}$, the $t-1$ prior outcomes of the pool-tests, the previous feedback symbols, $F^{t-1}$, and some randomness.
That is, we assume the lab may use a distribution $p(F_{t}|Y^{t-1},F^{t-1})$.
Hence, in the secure AGT \ifveryshort $X_{j}(t)= X_{j}(t|F^t)$. \else
\[
X_{j}(t)= X_{j}(t|F^t).
\]
\fi
Note that in the classical adaptive case, where simply the previous outcome is revealed to the mixer, we have $F_t=Y_{t-1}$, hence $|F_t|=2$ for all $t$ and therefor $R_f=1$.
\off{Hence, the symbol $F_{t}$ may depend on $Y^{t-1}$ and $F^{t-1}$.}

We assume the eavesdropper observes a noisy vector $Z^{T}=\{Z(1),\ldots, Z(T)\}$, generated from the outcome vector $Y^{T}$. We concentrate on the erasure case, where the probability of erasure is $1-\delta$, i.i.d.\ for each test.
That is, on average, $T\delta$ outcomes are not erased and are available to the eavesdropper via $Z^{T}$.

Denote by $W \in \mathcal{W} \triangleq \{1,\ldots, {N \choose K}\}$ the index of the \emph{subset of defective items}. We assume $W$ is uniformly distributed, that is, there is no \emph{apriori} bias to any specific subset. Denote by $\hat{W}(Y^T, F^T)$ the index recovered by the legitimate decoder, after observing $Y^T$. We refer to the adaptive procedure of creating the testing matrix, together with the decoder as an \emph{SAGT algorithm}.
We are interested in the asymptotic behavior of such algorithms.
\begin{definition}\label{def. security}
A sequence of SAGT algorithms with parameters $N,K$,$T$ and $R_f$ is asymptotically \emph{reliable} and \emph{weakly secure} if:

\noindent (1) At the legitimate receiver, observing $Y^T$, \ifshort $\lim_{T \to \infty} P(\hat{W}(Y^T, F^T) \ne W) = 0$. \else
\begin{equation*}
\lim_{T \to \infty} P(\hat{W}(Y^T, F^T) \ne W) = 0.
\end{equation*}
\fi
(2) At the eavesdropper, observing $Z^T$, \ifshort $\lim_{T \to \infty} \frac{1}{T}I(W;Z^T) = 0$. \else
\begin{equation*}
\lim_{T \to \infty} \frac{1}{T}I(W;Z^T) = 0.
\end{equation*}
\fi
\end{definition}
To conclude, The goal is to design (for parameters $N$, $K$ and $R_f$) an adaptive measurement scheme, where a feedback mechanism which depends on the outcome of previous tests controls the new tests, and a decoding algorithm, such that, observing $Y^{T}$ and $F^T$, the legitimate lab will identify the subset of defective items (with high probability), yet, observing $Z^{T}$, the eavesdropper cannot (asymptotically) infer the status of the items.
\subsubsection*{Batch Processing} The problem definition above aims at identifying the relation between $K,N,\delta,R_f$ and $T$ such that secure adaptive group testing is possible in the context of \Cref{def. security}. However, to practically accomplish the above goal, we will consider \emph{batch-processed} GT. In fact, this is the common method in practice for COVID-19 \cite{shental2020efficient}. Specifically, we will consider $B$ batches of $N$ items each, each batch containing $K$ defectives. The parameters $\delta$ and $R_f$ remain, that is, Eve observes all tests performed, in all batches, through a BEC($1-\delta$), and a feedback at a rate of $R_f$ bits per test is available. As will be proved in the sequel, such a batch processing scheme will be able to efficiently \emph{attain the lower bound we derive, valid for any scheme of adaptive and secure group testing}. Finally, \Cref{LowerBound_moel2} includes an enhancement of the main scheme, which does not use batches and employs a test-by-test feedback process, on a single batch.                                %
\section{Related Work}\label{BooleanCompressed}
\off{A GT model was first introduced in a landmark work \cite{dorfman1943detection} by Dorfman in the $1940s$.
This model was considered in many works as a combinatorial problem, where the papers provided efficient algorithms and bounds on the minimum number of group tests required.
One setting is where the entire pooling strategy is decided on beforehand, called \emph{Non-Adaptive GT} \cite{du1993combinatorial,chan2014non}. The other is where the outcomes of previous tests can be used to influence the makeup of future pools, called \emph{Adaptive GT} \cite{muller2001adaptive}. Anther aspect of the problem is whether the result of each test is noiseless or noisy, namely, where possible errors occur in the pool testing \cite{atia2012boolean}.

The bounds available in the literature on the number of tests required to identify all the defective items in noiseless NGT, are $T=\Theta (K\log N)$. See, e.g., \cite{atia2012boolean}. In fact, it is exactly $K\log N$ in the simple Boolean scenario.
The bounds on the number of tests required in the noisy scenario are, in general, a constant factor larger than the noiseless scenario \cite{chan2011non}.}
Recently works \ifveryshort\cite{atia2012boolean, aksoylar2014information, cohen2016secure, aldridge2012adaptive, baldassini2013capacity, aeron2010information, kealy2014capacity} \else\cite{atia2012boolean, aksoylar2014information, cohen2016secure, aldridge2012adaptive, malyutov2013search, aydinian2013information, baldassini2013capacity, aeron2010information, kealy2014capacity,aldridge2017almost,coja2019information,aldridge2019group,coja2019optimal,bshouty2020optimal,tan2020near} \fi adopted an information theoretic perspective on GT, presenting it as a channel coding problem.
We briefly review the most relevant results.

In \cite{atia2012boolean}, the authors mapped the NGT model to an equivalent channel model, where the defective set takes the place of the message, the testing matrix rows are codewords, and the test outcomes are the received signal.
They let $\hat{S}(X^T, Y^T)$ denote the estimate of the defective subset $S$, which is random due to the randomness in $X$ and $Y$. Furthermore, let $P_e$ denote the probability of error, averaged over all subsets $S$ of cardinality $K$, variables $X^T$ and outcomes $Y^T$, i.e., $P_e=Pr[\hat{S}(X^T, Y^T)\neq S]$.
Then, if $(\mathcal{S}^{1},\mathcal{S}^{2})$ denotes the partition of the defective set $S$ into disjoint sets $\mathcal{S}^{1}$ and $\mathcal{S}^{2}$ with cardinalities $i$ and $K-i$, respectively, and replacing $\log\binom{N-K}{i}$ by it's asymptotic value $i\log N$ (when $K \ll N$) , the flowing bounds on the total number of tests required in Bernoulli NGT were given: $\underline{T}\leq T \leq \overline{T}$, where for some $\varepsilon > 0$ independent of $N$ and $K$, \ifveryshort $\overline{T} = (1+\varepsilon)\cdot\max_{i=1,\ldots ,K} \frac{\log\binom{N-K}{i}}{I(X_{\mathcal{S}^1};X_{\mathcal{S}^2},Y)}$ and $\underline{T} = \max_{i=1,\ldots ,K} \frac{\log\binom{N-K+i}{i}}{I(X_{\mathcal{S}^1};X_{\mathcal{S}^2},Y)}$. \else
\begin{equation*}\label{T_ns_na_hi}
  \overline{T} \triangleq (1+\varepsilon) K \log N ,
\end{equation*}
\begin{equation*}\label{T_ns_na_lo}
  \underline{T} \triangleq \log\binom{N}{K}.
\end{equation*}
\fi

In \cite{aldridge2012adaptive}, the authors  considered the AGT model as channel coding with feedback, where future inputs to the channel can depend on past outputs.
Shannon proved that feedback does not improve the capacity of a single-user channel \cite{shannon1956zero}.
However, due to the non-tightness of the bounds on testing in the non-adaptive case, the authors could not show that adaptive GT requires the same number of tests as non-adaptive testing exactly. Alternatively, they showed that it obeys the same lower bound and requires no more tests than the non-adaptive case.
\off{
In AGT, the choice of a testing pool depend on the outcomes of earlier tests.
This is similar to channel coding with feedback, where future inputs to the channel can depend on past outputs.
Shannon proved that feedback does not improve the capacity of a single-user channel \cite{shannon1956zero}.
However, in \cite{aldridge2012adaptive} showed proceeds similarly, that due to the non-tightness of the bounds on testing in the non-adaptive case, they couldn't able to show that adaptive group testing requires the same number of tests as non-adaptive testing, but they showed that it obeys the same lower bound and requires no more tests than the non-adaptive case.}
\ifveryshort

\else
\off{
\begin{theorem}[2,\cite{aldridge2012adaptive}]\label{theorem_atia}
Let $N$ be the size of a population of items with the defective set $S$ of cardinality $K$.
Where, $T$ and $T^{a}$ be the minimum number of tests necessary in the non-adaptive model and adaptive model respectively, we have the inequalities \ifshort $\underline{T}\leq T^{a}\leq T\leq \overline{T}$ \else
\begin{equation*}
\underline{T}\leq T^{a}\leq T\leq \overline{T}
\end{equation*}
\fi
where $\underline{T}$ and $\overline{T}$ are as in \eqref{T_ns_na_hi} and \eqref{T_ns_na_lo},
then asymptotically the average error probability approaches zero for adaptive and non-adaptive models, namely, \ifshort $\lim_{K\rightarrow \infty} \lim_{N\rightarrow \infty} P_e \rightarrow 0$ \else
\begin{equation*}
\lim_{K\rightarrow \infty} \lim_{N\rightarrow \infty} P_e \rightarrow 0
\end{equation*}
\fi
where $\varepsilon > 0$ is a constant independent of $N$ and $K$.
\end{theorem}}

\ifshort\else
In \cite{aydinian2013information}, an optimal AGT scheme was proposed. This scheme involves several testing stages, using, in each stage, the results of the previous stages. Moreover, in \cite{baldassini2013capacity}, using a similar analogy to the channel coding problem, the authors defined the \emph{AGT rate} $R$ as $\log_2 \binom{N}{K}/T$ and introduced the capacity $C$, that is, a constant such that for any $\varepsilon>0$, there exists a sequence of algorithms with \ifshort $\lim_{N\rightarrow \infty} \frac{\log_2\binom{N}{K}}{T^{a}}\leq C - \varepsilon$ \else
\begin{equation*}
\lim_{N\rightarrow \infty} \frac{\log_2\binom{N}{K}}{T}\leq C - \varepsilon
\end{equation*}
\fi
and success probability approaching one, yet if $\lim_{N\rightarrow \infty} \log_2\binom{N}{K}/T > C$, it approaches zero.
\fi
\fi
Conducing the results above, for both adaptive and non-adaptive GT, the capacity is $1$.
Recently AGT were considered for coronavirus disease 2019 (COVID-19) in \cite{narayanan2020accelerated,theagarajan2020group}.

In our earlier work \cite{cohen2016secure}, we focused on SNGT.
We considered a scenario where there is an eavesdropper which is able to observe a subset of the outcomes.
We proposed a SNGT algorithm, which keeps the eavesdropper with leakage probability $\delta$, ignorant regarding the items' status. \ifshort \else Specifically, when the fraction of tests observed by Eve is $0 \leq \delta <1$, \cite{cohen2016secure} proved that the number of tests required for both correct reconstruction at the legitimate user and negligible mutual information at Eve's side is $\frac{1}{1-\delta}$ times the number of tests required with no secrecy constraint. \fi
Then, \off{where $I(X_{\mathcal{S}^1};X_{\mathcal{S}^2},Y)\geq i/K$, according \cite[Claim 1]{cohen2016secure}, the }flowing bounds on the total number of tests required in Bernoulli SNGT was given by  $\underline{T}_{s}\leq T \leq \overline{T}_{s}$, where for some $\varepsilon > 0$ independent of $N$ and $K$, \ifveryshort $\overline{T}_{s} = \frac{1+\varepsilon}{1-\delta} \max_{i=1,\ldots ,K} \frac{K}{i}\log\binom{N-K}{i}$ and $\underline{T}_{s} = \frac{1}{1-\delta}\log\binom{N}{K}$.\else
\ifshort
\begin{equation}\label{T_s_na_hi}
 \overline{T}_{s} = \frac{1+\varepsilon}{1-\delta} \max_{i=1,\ldots ,K} \frac{K}{i}\log\binom{N-K}{i}
\end{equation}
\begin{equation}\label{T_s_na_lo}
  \underline{T}_{s} = \frac{1}{1-\delta}\log\binom{N}{K}
\end{equation}
\else
\begin{equation}\label{T_s_na_hi}
 \overline{T}_{s} \triangleq \frac{1 + \varepsilon}{1 - \delta} K \log N ,
\end{equation}
\begin{equation*}\label{T_s_na_lo}
 \underline{T}_{s} \triangleq \frac{1}{1-\delta}\log\binom{N}{K} .
\end{equation*}
\fi
\fi                                 %
\section{Main Results}\label{main results}
Under the model definition given in \Cref{formulation}, our main results are the following sufficiency (direct) and necessity (converse) conditions, characterizing the number of tests required to guarantee both reliability and security. 

\subsection{Direct (Sufficiency)}
With a private rate-limited feedback link, for $B$ batches of $T$ tests each, the sufficiency part is given by the following theorem.
\begin{theorem}\label{direct theorem1}
Assume a SAGT model with $N$ items, out of which $K=O(1)$ are defective. For any $\delta < 1$ and private adaptive rate-limited feedback $0\leq R_f\leq 1$, if
\begin{equation*}\label{main_result_eq}
   T \geq \frac{1+\varepsilon}{\min\{1,1-\delta+R_f\}} K \log N  \triangleq \overline{T}_{s}^{a},
\end{equation*}
for some $\varepsilon > 0$ independent of $N$ and $K$, then there exists a sequence of SAGT algorithms which are reliable and weakly secure. That is, as $N\rightarrow \infty$, both the average error probability approaches zero and an eavesdropper with leakage probability $\delta$ is kept ignorant.
\end{theorem}
The construction of the SAGT algorithm with private rate-limited feedback, together with the proofs of reliability and secrecy are deferred to Section \ref{LowerBound}.

At this point, a few remarks are in order. First, it is important to note that if $R_f \geq \delta$, as the direct proof will show, the information obtained over the adaptive link between the lab and the mixer is powerful enough to obtain security without increasing $T$ compared to the non-secure case. Hence, in this case, the direct bounds of the non-secure and secure adaptive group testing are equal, that is, $\overline{T}=\overline{T}^{a}_{s}$.
Even when $ R_f < \delta$, the information obtained over the feedback between the lab and the mixer reduces the upper bound on the number of the tests required in SNGT, thus, $\overline{T}<\overline{T}^{a}_{s}\leq\overline{T}_{s}$.

Second, \Cref{direct theorem1} refers to the asymptotic case in $T$, and negligible error. In this case, without a secrecy constrain, it is well known that feedback does not help \cite{aldridge2012adaptive}. With the secrecy constraint, we show that the link is \emph{optimally used} only to obtain shared randomness.
However, for finite $T$ and zero error \cite{du2000combinatorial}, the mixer may use the link differently, including sending previous outcomes. Thus, for finite $T$, zero error and a secrecy constraint, it is not trivial what should be shared over the link: either pure randomness (which is optimal in the asymptotic scenario) in order to cope with the secrecy constraint or previous outcomes in order to try adaptively reduce the number of tests.

Third, to simplify the technical aspects and focus on the key methods, in this paper, we use Maximum Likelihood decoding to obtain the sufficiency result in \Cref{main_result_eq}, where $K=O(1)$. However, the same techniques proposed in this paper for adaptive GT using the private feedback link may be applied when using computationally efficient decoding algorithms proposed for GT in the literature, at a price of slightly more number of tests \cite{kautz1964nonrandom,chen2008survey,chan2011non,chan2014non,aldridge2014group,lo2013efficient,malyutov2013search,aldridge2019group,cohen2016secure}.\off{ For instance, considering each codeword in the codebook proposed in \Cref{LowerBound} as an independent item.} In particular, using the same codebook and key generation we suggest herein, and the testing as given in \Cref{LowerBound}, but decoding with the \emph{Definitely Non-Defective} algorithm proposed in the literature \cite{kautz1964nonrandom}, computationally efficient decoding and secrecy are possible, at the price of a higher $T$ for any $K$, and not necessarily only for $K=O(1)$.  

\off{First, rearranging terms in \cref{main_result_eq}, we have
\begin{equation*}\label{main_result}
 T_{s}^{a}  \geq   \frac{1}{\min\{1,1-\delta+R_f\}} \max_{i=1,\ldots ,K}\frac{(1+\varepsilon)K}{i}\log\binom{N-K}{i}.
\end{equation*}
That is, compared to only a reliability constraint, the number of tests required for \emph{both reliability and secrecy} is increased by the multiplicative factor $\frac{1}{\min\{1,1-\delta+R_f\}}$, where, again, $\delta$ is the leakage probability at the eavesdropper and $R_f$ is the adaptive private rete limited feedback from the legitimate user.

Since for a fixed $K$ and large enough $N$ we have $\log\binom{N-K}{i} = \Theta(i\log N)$, we have the following corollary on the number of tests required.
\begin{corollary}
For SAGT with parameters $K << N$, private rete limited feedback $R_f$ and $T$, reliability and weak secrecy can be maintained with
\begin{equation*}
 T_{s}^{a} = \Theta \left(\frac{K\log N}{\min\{1,1-\delta+R_f\}}\right).
\end{equation*}
\end{corollary}
Note that this suggests a $\Theta \left(K\log N\right)$ result for $\delta$ bounded away from $1$.

The results thus far were for $K=O(1)$. Clearly it is interesting to analyse the regime where $K$ is allowed to grow with $N$, though at a smaller rate, that is $K=o(N)$.}
\subsection{Converse (Necessity)}
The necessity part is given in the following theorem.
\begin{theorem}\label{converse theorem}
The minimum number of tests necessary to identify a defective set of cardinality $K$ among a population of size $N$ in an SAGT model with private adaptive rate limited feedback $0\leq R_f\leq 1,$ while keeping eavesdropper, with pooling outcome test leakage probability $\delta < 1$, ignorant regarding the status of the items is
\begin{eqnarray*}\label{T_s_a_lo}
 T \ge  \frac{1}{\min\{1,1-\delta+R_{f}\}}\log\binom{N}{K} \triangleq  \underline{T}_{s}^{a},
\end{eqnarray*}
such that $\frac{1}{T}I(W;Z^{T}) \leq \epsilon_T + \upsilon_T$ where $\epsilon_T \to 0$ and $\upsilon_T \to 0$ as $T \to \infty$.
\end{theorem}

\ifshort
Due to space limitation, the proof is deferred to the extended version of this paper \cite[Section VII]{cohen2018secure}.
\else
The proof is deferred to Section \ref{converse}.
\fi
Note that, compared to the lower bound without a security constraints, there is an increase by a multiplicative factor of $1/\min\{1,1-\delta+R_f\}$. When $R_f \geq \delta$, the lower bounds of the non-secure and secure adaptive group testing are equal, i.e., $\underline{T}=\underline{T}^{a}_{s}$.
\ifshort\else
\subsection{Secrecy capacity in SAGT}
Returning to the analogy in \cite{baldassini2013capacity} between channel capacity and group testing, one might define by $C_s$ the (asymptotic) minimal threshold value for $\log\binom{N}{K}/T$, above which no reliable and secure scheme is possible. Under this definition, where $C$ is the capacity without the security constraint, the result in this paper show that $C_s \ge (\min\{1,1-\delta+R_f\})C $
Clearly, this can be written as, \ifshort $C_s \ge \min\{C, C - C(\delta)+C(R_f)\}$, \else
\begin{eqnarray*}
C_s &\ge& \min\{C, C - C \delta + CR_f\},
\end{eqnarray*}
\fi
raising the usual interpretation as the \emph{difference} between the capacity to the legitimate decoder and that to the eavesdropper, yet, while adding the rate of information obtained over the feedback link, between the legitimate decoder and encoder \cite{ahlswede2006transmission, C8, cohen2016wiretap}. This is since the effective number of tests Eve sees is $\delta T$, hence her GT capacity is $\delta C$.
\fi 

\subsection{SAGT using curious but non-malicious non-cooperative labs}
We further consider two additional models of SAGT using curious but non-malicious non-cooperative labs (\Cref{figure:group_testing_model_doctor}). One with a private feedback link from the doctor and the second with direct feedback links from the labs (hence known to the labs sending them). In both models, a nurse Alice collects the blood samples from the patients and prepares tubs of pool-tests using a testing matrix. Then she chooses $L$ independent and curious but non-malicious non-cooperative labs, and sends each lab a set of tests. Let $\delta$ denote the maximum fraction of the pool-test a lab gets to test (for simplicity, assume $\delta \leq 1/2$ and $L=1/\delta$ an integer). The test results from all the labs are provided to doctor Bob to be analyzed. The goal is that only Bob will identify the subset of the defective items, yet, each lab observing only one set of tubes cannot infer any information about the patients.

When the doctor's clinic can send the nurse information at a certain rate $R_f$ over private feedback link, we show in \Cref{sec:labs} that the sufficiency part is given by \Cref{direct theorem1} easily. However, when the private feedback link is not available, yet each independent lab can send over a direct link feedback to the nurse, \emph{hence its content is available to that curious lab}, an analogous result still holds, and we have the following.  
\begin{theorem}\label{direct_bab}
Assume a SAGT model with $N$ items, out of which $K=O(1)$ are defective. Assume $\delta \leq 1/2$ and chosen $L = 1/\delta$ curious but non-malicious non-cooperative labs, each receiving a $\delta$ fraction of the tests and having a direct link at rate $R_{pf}$. Reliability and secrecy can be maintained if
\begin{equation*}\label{main_result_lab}
  T \geq \frac{1+\varepsilon}{\min\{1,1-\delta+R_{pf}\}} K \log N = \overline{T}_{s}^{a},
\end{equation*}
for some $\varepsilon > 0$ independent of $N$ and $K$.
\end{theorem}
The details are deferred to \Cref{sec:labs}. It is important to note here, however, that a key contribution herein is that the random key using in this scheme is obtained using \emph{solely the information from the direct link, which is known to the curious lab}. This key (for $R_f = \delta$) is sufficient to obtain security without increasing $T$. That is, in this case $\overline{T}=\overline{T}^{a}_{s}$.
                                %
\section{Code Construction with Adaptive Private Feedback\\ and a Proof for \Cref{direct theorem1}} \label{LowerBound}
The goal, in general, is to design a proper testing matrix, or, specifically, an algorithm to adaptively update it. Remember that each row describes the tests an item participates in. Security is achieved by randomization: selecting a row at random for each item. However, the goal is to properly design the sets of rows to choose from, and intelligently use both the feedback and internal randomization, such that the eavesdropper is confused due to the numerous rows possible, yet the legitimate receiver, with its knowledge of the feedback bits, can decode.

\begin{figure}
  \centering
  \ifjsait
  \includegraphics[trim= 0cm 0cm 0cm 0cm,clip,scale=1.2]{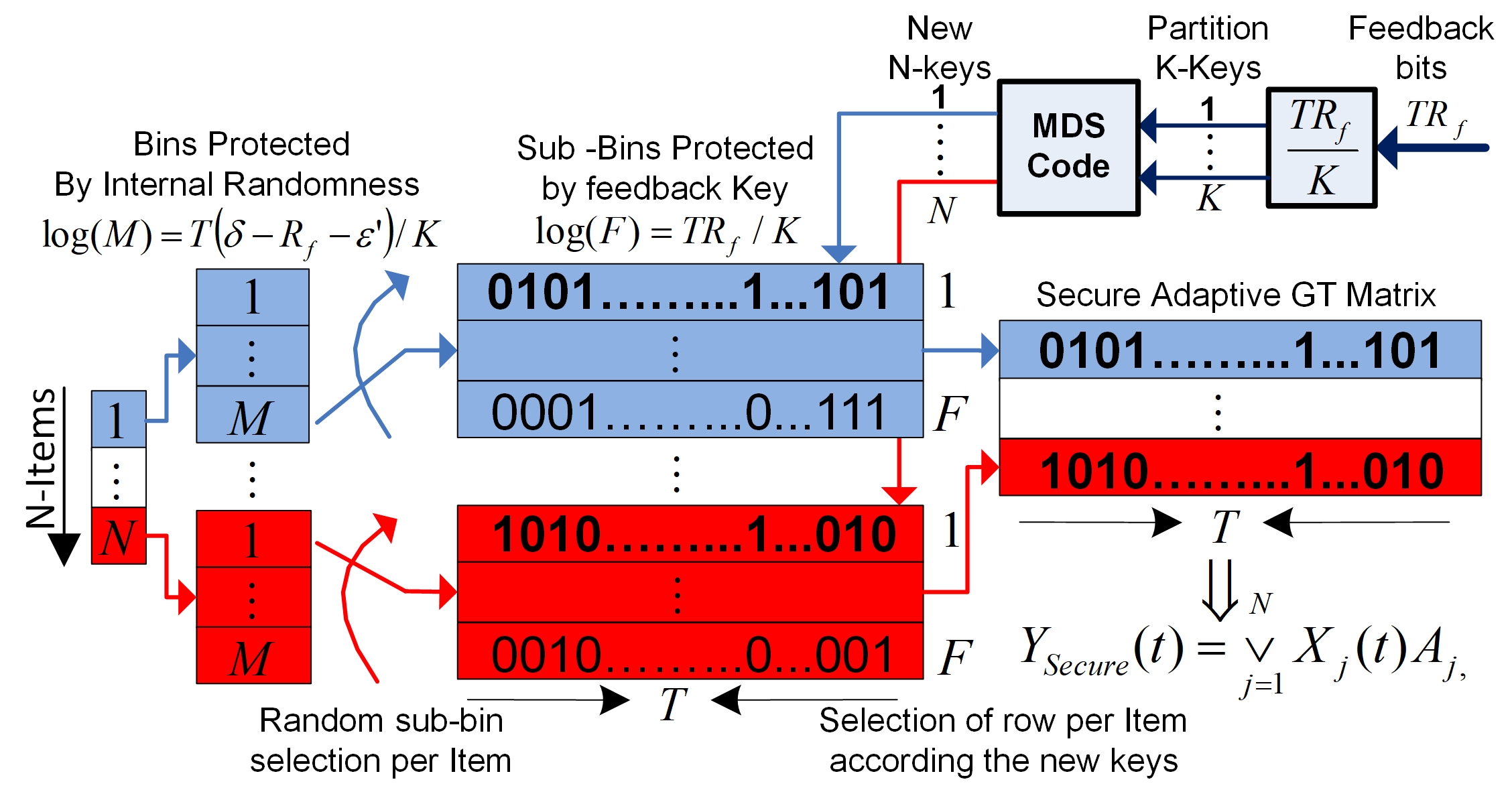}
  \else
  \includegraphics[trim= 0cm 0cm 0cm 0cm,clip,scale=0.95]{fig/Wiretap_coding5_one_col_AdpS2.jpg}
  \fi
  \caption{Testing process for a SAGT testing matrix.}
  \label{fig:WiretapCoding}
\end{figure}

We thus construct this matrix in $B$ batches (of T tests each), each time selecting an appropriate row for each item. In a batch $b\in\{1,\ldots,B\}$, for each item we have a bin, containing several sub-bins, with several rows in each sub-bin (see \Cref{fig:WiretapCoding}). Internal randomness in the mixer, which is not shared with any other party, is used to select a sub-bin for each item, while data received from the adaptive link is used to select a row from the sub-bin. While this solution is inspired by codes for wiretap channels with rate-limited feedback \cite{C8}, there are several key differences, which not only change the construction, but also require \emph{non-trivial processing of the data received from the feedback}. Specifically, first, unlike a wiretap channel, herein there are $N$ items, yet only $K$ of them, unknown to the mixer, actually participate in the output (``transmit''). Thus, bins and sub-bins sizes should be properly normalized. More importantly, the mixer, which acts as an encoder, \emph{does not know which $K$ messages it should protect}. Thus, the mixer should artificially blow-up the data it receives from the  private feedback: from bits (used as keys) intended to protect the $K$ defective items, it generates a larger number of keys, sufficient for $N$ items, and satisfying the property that any $K$ out of the $N$ which will eventually participate, will still be protected. In other words, the keys received from the feedback cannot be used as is, and an interesting \emph{secret sharing-type} scheme must be used.

Formally, a (batch-processing) SAGT code consists of an index set $\mathcal{W} =\{1,2,\ldots \binom{N}{K}\}$, its $w$-th item corresponding to the $w$-th subset $\mathcal{K}\subset \{1,\ldots,N\}$; A discrete memoryless source of randomness at the mixer $(\mathcal{R_X},p_{R_X})$; A discrete memoryless source of randomness at the lab $(\mathcal{R_Y},p_{R_Y})$; A feedback at rate $R_f$ bits per test, resulting in an index $\mathcal{I}\in \{1,\ldots,2^{TR_f}\}$ after $T$ uses; We use a single index due to the batch processing. \Cref{LowerBound_moel2} describes a test-by-test adaptive algorithm based on the one herein. The mixer, of course, does not know which items are defective, thus it needs to select a row for \textit{each} item. However, since only the rows of the defective items affect the output $Y^T$, it will be beneficial to define an ``encoder" \ifshort $\mathcal{G_X} : \mathcal{W} \times \mathcal{R_X} \times \mathcal{I} \rightarrow \mathcal{X}_{S_w}\in\{0,1\}^{K\times T},$ \else
\begin{equation*}
\mathcal{G_X} : \mathcal{W} \times \mathcal{R_X} \times \mathcal{I} \rightarrow \mathcal{X}_{S_w}\in\{0,1\}^{K\times T},
\end{equation*}
\fi
that is, mapping the actual index of the defective items, the randomness and input from the adaptive link to a testing matrix $X_{S_{w}}^{T}$ of $K$ codewords, which are summed to give the tests output $Y^T$. The codewords in the testing matrix correspond to the index set of the defective items $S_{w}$, where $|S_{w}|=K$. Note that a \emph{stochastic encoder} and the causally known feedback message $\mathcal{I}$ are similar to encoders ensuring information theoretic security, as randomness is required to confuse the eavesdropper about the actual information \cite{C13,C8}.

A decoder at the legitimate lab is a map \ifshort $\hat{W} : Y^{T} \times \mathcal{I} \rightarrow \mathcal{W}$. \else
\begin{equation*}
\hat{W} : Y^{T} \times \mathcal{I} \rightarrow \mathcal{W}.
\end{equation*}
\fi
The probability of error is $P(\hat{W}\neq W)$. The probability that an outcome test leaks to the eavesdropper is $\delta$. We assume a memoryless model, i.e., each outcome $Y(t)$ depends only on the corresponding input $X_{S_w}(t)$, and the eavesdropper observes $Z(t)$, generated from $Y(t)$ according to \ifshort $p(Y^T,Z^T|X_{S_w})=\prod_{t=1}^{T}p(Y(t)|X_{S_w}(t))p(Z(t)|Y(t))$. \else
\begin{equation*}
p(Y^T,Z^T|X_{S_w})=\prod_{t=1}^{T}p(Y(t)|X_{S_w}(t))p(Z(t)|Y(t)).
\end{equation*}
\fi
Next, we provide the detailed construction and analysis.

\subsubsection{Codebook Generation}\label{LowerBound1}
For $b\in\{2,\ldots,B\}$, fix $T$ satisfying the condition in \Cref{direct theorem1}. Choose integers $F$ and $M$ such that \ifshort $\log_2(F) = T(R_f/K) \mbox{ and } \log_2(M) = T(\delta-R_f-\epsilon)/K.$ \else $$\log_2(F) = T(R_f/K) \mbox{ and } \log_2(M) = T(\delta-R_f-\epsilon)/K.$$ \fi
Note we assume $R_f < \delta$. As will be clear in the sequeal, if $R_f \ge \delta$ there is no need for two binning layers. We fix $M=1$ and $\log_2(F) = T(\delta-\epsilon)/K$, namely, sub-bins protected by the feedback key will suffice.
For each item we generate bin with $M\cdot F$ independent and identically distributed binary codewords. Each codeword of size $T$ is generated randomly, where each bit is chosen according to a $Bernoulli(\ln(2)/K)$ distribution.
Then, we split each bin to $M$ sub-bins of $F$ codewords each.  $x^{T}(m,f)$, $1 \leq m \leq M$ and $1 \leq f \leq F$ denotes a codeword (Figure~\ref{fig:WiretapCoding}). Denote by $\mathcal{C}_{1}$ the random codebook used for batches $b\in\{2,\ldots,B\}$.

For the first batch, since the mixer has no available key from the feedback, fix $T$ satisfying the condition in \eqref{T_s_na_hi}, and create a codebook as given above when $\log_2(F)=1$ and $\log_2(M)=T(\delta-\epsilon)/K$. This results in a second codebook $\mathcal{C}_{2}$ with one codeword in each sub-bin.
\subsubsection{Key Generation}\label{LowerBound2}
We now describe the generation of the shared keys created from the information sent over the adaptive link. This link is of rate $R_f$. \emph{We do not use it to send information about test results, and simply send random bits}. Therefore, in a block of length $T$ we receive $S=TR_f$ secret bits. We divide the secret bits to $K$ secret keys, each constituting $TR_f/K=S_K$ bits.

Our goal is to take these $K$ keys, and use them to create $N$ new keys, with the property that if the original $K$ keys had a random uniform i.i.d. distribution, then \emph{any set of $K$ keys out the $N$ new ones} will have the same random uniform i.i.d. distribution. This can be done using a generator matrix of an $[N,K]$ MDS code\footnote{For example, Reed Solomon and Rank-Metric Codes given in \cite{wicker1999reed} and \cite{silva2011universal}, respectively, can be used for the generator matrix.}. Such a generator matrix has the property that any $K$ columns are linearly independent. Thus, taking the $K$ original keys, as an $S_K\times K$ matrix, and multiplying it by the generator matrix $G_{K\times N}$ creates a matrix of size $S_K \times N$, where each column is used as the new key. Since any subset of $K$ columns of $G$ is invertible, each set of $K$ new keys is simply a $1:1$ transformation of the $K$ original keys. The importance of this scheme in our context is as follows: for any subset of $K$ new keys (out of $N$), if an eavesdropper has no access to the original $K$ keys, he/she is ignorant regarding the new keys (in a sense to be made precise in \Cref{leakage}). Note also that unlike protection using a one-time-pad, usually done when the two legitimate parties have a shared key, these keys cannot be XORed with the rows of the testing matrix, as this operation will change the probability of each item to participate in a pool-test, deviating from the optimal distribution of the testing matrix.
\subsubsection{Testing}\label{LowerBound3}
For $b\in\{2,\ldots,B\}$, The mixer receives the $TR_f$ feedback secret bits, divides them to $K$ keys of length $TR_f/K$ and uses the MDS code to create $N$ new keys. Each new key is of length $TR_f/K$ as well. Therefore, using $\mathcal{C}_{1}$ at each round and for each item $j$, the mixer selects a sub-bin using its internal randomness, and a message within it using the key. This codeword, $x^{T}(m,f)$, will constitute the $j$' th row in the testing matrix of the current batch. Therefore, the SAGT matrix contains $N$ randomly selected codewords of length $T$, one for each item (defective or not).

For the first batch of tests, the mixer has no feedback key, hence the mixer operates using the second codebook $\mathcal{C}_{2}$. The mixer selects a sub-bin, with one codeword, using its internal randomness. This codeword, $x^{T}(m,1)$, will constitute the $j$' th row in the testing matrix of the first batch. Therefore, the testing matrix contains $N$ randomly selected codewords of length $T$ satisfying the condition in \eqref{T_s_na_hi}. Hence, the first batch is longer by a constant multiplier, yet amortized over multiple rounds, the loss in number of pool-tests required in the first batch is negligible.
\subsubsection{Decoding at the Legitimate Receiver}\label{LowerBound4}
For a batch $b\in\{2,\ldots,B\}$, the decoder uses the feedback bits sent to the mixer at the end of the previous batch, in order to pick the appropriate codeword from each sub-bin. Then, the decoder looks for a collection of $K$ bins, each containing a single codeword, and finds the resulting set of $K$ codewords, $X_{S_{\hat{w}}}^{T}$, for which $Y^T$ is most likely. Namely, \ifshort $P(Y^{T}|X_{S_{\hat{w}}}^{T})>P(Y^{T}|X_{S_{w}}^{T}), \forall w \neq \hat{w}$. \else
\begin{equation*}
P(Y^{T}|X_{S_{\hat{w}}}^{T})>P(Y^{T}|X_{S_{w}}^{T}), \forall w \neq \hat{w}.
\end{equation*}
\fi
The legitimate lab declares $\hat{W}(Y^T \times F)$ as the set of bins in which the rows reside. For the first batch, as there is no key to narrow down the number of codewords, the decoder performs the same test, yet on the entire codebook $\mathcal{C}_{2}$. Remember that this codebook has a larger number of codewords, yet each is longer.
\subsubsection{Reliability}
\off{Recall $(\mathcal{S}^{1},\mathcal{S}^{2})$ denote the partition of defective set $S$ into disjoint sets $\mathcal{S}^{1}$ and $\mathcal{S}^{2}$ with cardinalities $i$ and $K-i$, respectively. Let $I(X_{\mathcal{S}^1};X_{\mathcal{S}^2},Y)$ denote the mutual information between $X_{\mathcal{S}^1}$ and $(X_{\mathcal{S}^2},Y)$, under the i.i.d.\ distribution with which the codebook was generated and remembering that $Y$ is the output of a Boolean channel.}

The following lemma is a key step in proving the reliability of the decoding algorithm suggested herein. This Lemma is a direct consequence of \cite[Lemma 1]{cohen2016secure} applying \cite[Claim 1]{cohen2016secure}, under the enhancement that the index of the row in each sub-bin is set according to the known key sheared between the legitimate decoder and the mixer at each round of the algorithm.
\begin{lemma}\label{direct lemma1}
If for some $\varepsilon$ independent of $K$ and $N$ the number of tests satisfies
\[
T  \geq (1+\varepsilon)\cdot\max_{i=1,\ldots ,K} \frac{\log \left( \binom{N-K}{i}M^i \right) }{i/K},
\]
then, under the codebook above, as $N\rightarrow \infty$ the average error probability approaches zero.
\end{lemma}
\off{Applying \cite[Claim 1]{cohen2016secure}, which lower bounds the mutual information between $X_{\mathcal{S}^1}$ to $(X_{\mathcal{S}^2},Y)$ by $i/K$, to the expression in \Cref{direct lemma1}, \off{we have
\begin{equation}\label{ApplyingClaim}
\frac{\log\binom{N-K}{i}M^i}{I(X_{\mathcal{S}^1};X_{\mathcal{S}^2},Y)} \leq \frac{\log\binom{N-K}{i}M^i}{\frac{i}{K}}.
\end{equation}
Hence,} and} First, note that the paramter $F$ of the codebook, the number of rows within a sub-bin is irrelevant, as the row within the sub-bin is known to both legitimate parties. Now, substituting $M=2^{T\frac{\delta-R_f-\epsilon }{K}}$ in the expression in \Cref{direct lemma1}, a sufficient condition for reliability is
\begin{eqnarray*}
T &\ge& \max_{1 \leq i \leq K}\frac{1+\varepsilon}{\frac{i}{K}}\left[ \log\binom{N-K}{i} + \frac{i}{K}T(\delta-R_f) \right].
\end{eqnarray*}
Rearranging terms and remembering that for $R_f \ge \delta$ we take $M=1$ results in
\begin{eqnarray}\label{eq:reduce_h}
\hspace{-0.02cm} T &\hspace{-0.3cm} \ge & \hspace{-0.3cm} \max_{1 \leq i \leq K} \frac{1+\varepsilon}{\min\{1,1-(1+\varepsilon)(\delta-R_f)\}i/K}\log\binom{N-K}{i}.\nonumber
\end{eqnarray}
Now, replacing $\log\binom{N-K}{i}$ by it's asymptotic value $i\log N$ (for $K=O(1)$), the maximization can be solved easily and we have
\begin{equation*}
 T \geq  \frac{1+\varepsilon}{\min\{1,1-\delta+R_f\}} K \log N.
\end{equation*}
This complete the reliability part.
\subsubsection{Information Leakage at the Eavesdropper}\label{leakage}
We now prove the security constraint is met. Recall that the codewords length $T$, for the first batch and the rest, satisfying the conditions in \eqref{T_s_na_hi} and \Cref{direct theorem1}, respectively. Yet, according to \Cref{def. security}, security is judged based on the entire set of tests. Hence, for $B$ batches we wish to show that $I(W;Z^{BT})/BT\rightarrow 0$, as $T\rightarrow \infty$. Denote for each batch $b\in\{1,\ldots,B\}$, by $X_{\mathcal{S},b}^T$ the set of codewords corresponding to the true defective items $W_b$ and by $Z^{T}_{b}$ the eavesdropper's observation in batch $b$. Hence, the information leaked, given the random codebook $\mathcal{C} = (\mathcal{C}_{1}, \mathcal{C}_{2})$ is
\begin{equation*}
 \hspace{-0.3cm}\frac{1}{TB} I(W_1,\ldots, W_B;Z^{T}_{1},\ldots,Z^{T}_{B}|\mathcal{C})
 = \frac{1}{TB} \sum_{b=1}^{B} I(W_b;Z^{T}_{b}|\mathcal{C}),
\end{equation*}
where the equality follows since the feedback link includes only purely random bits, used only for a single batch of tests. Hence, $Z^{T}_{b}$, which depends on the codewords in batch $b$, is affected only by $W_b$ and random bits, parts of which are from the internal randomness at the mixer and parts of which were received through the feedback, yet all are independent of the inputs and outputs of other batches.

For each $b\in\{2,\ldots,B\}$ we have,
\ifjsait
\begin{equation}\label{eq:lekage}
 \frac{1}{T}I(W_b;Z^T_b|\mathcal{C}) = \frac{1}{T}\Big( I(W_b,R_{K,b}R_{K,b}^{F};Z^T_b|\mathcal{C}) \\
  -I(R_{K,b}R_{K,b}^{F};Z^T_b|W_b,\mathcal{C})\Big),
\end{equation}
\else
\begin{multline}\label{eq:lekage}
 \frac{1}{T}I(W_b;Z^T_b|\mathcal{C}) = \frac{1}{T}\Big( I(W_b,R_{K,b}R_{K,b}^{F};Z^T_b|\mathcal{C}) \\
  -I(R_{K,b}R_{K,b}^{F};Z^T_b|W_b,\mathcal{C})\Big),
\end{multline}
\fi
where $R_{K,b}$ is the internal randomness used by the encoder to choose the bins, while $R_{K,b}^{F}$ are the $K$ feedback keys. Note that in practice the encoder should have a random variable for \emph{each item}, and a feedback key \emph{for each item} (generated using the MDS code). However, for the leakage proof, we are interested only in the two sets of $K$ keys \emph{used for the defective items}. Thus, $R_{K,b}$ is the union of $K$ such internal randomness variables, and $R_{K,b}^{F}$ is again a union, this time of $K$ feedback keys - $K$ ``shares" out of the $N$ shares generated by the MDS code. Since $W_b,R_{K,b}$ and $R_{K,b}^{F}$ uniquely define $X_{\mathcal{S},b}^T$, continuing from \eqref{eq:lekage}, we have:
\ifjsait
\begin{eqnarray}
&  =  & \frac{1}{T}\left(I(X_{\mathcal{S},b}^{T};Z^T_b|\mathcal{C})-I(R_{K,b}R_{K,b}^{F};Z^T_b|W_b,\mathcal{C})\right)\nonumber\\
& = & \frac{1}{T}\Big(I(X_{\mathcal{S},b}^{T};Z^T_b|\mathcal{C})-H(R_{K,b}R_{K.b}^{F}|W_b,\mathcal{C}) + H(R_{K,b}R_{K,b}^{F}|Z^T_b,W_b,\mathcal{C})\Big)\nonumber\\
&  \stackrel{(a)}{=} &  \frac{1}{T}\Big(I(X_{\mathcal{S},b}^{T};Z^T_b|\mathcal{C}) -H(R_{K,b}R_{K,b}^{F})+H(R_{K,b}R_{K,b}^{F}|Z^T_b,W_b,\mathcal{C})\Big)\nonumber\\
& \stackrel{(b)}{\leq}& \delta - \frac{1}{T}K\left(T \frac{\delta-R_f-\epsilon}{K}+T\frac{R_f}{K}\right)+\frac{1}{T}H(R_{K,b}R_{K,b}^{F}|Z^T_b,W_b,\mathcal{C})\nonumber\\
& \stackrel{(c)}{\leq}&  \epsilon_{T}^{\prime\prime}(b),\nonumber
\end{eqnarray}
\else
\begin{eqnarray}
&\hspace{-0.6cm}  =  &\hspace{-0.5cm} \frac{1}{T}\left(I(X_{\mathcal{S},b}^{T};Z^T_b|\mathcal{C})-I(R_{K,b}R_{K,b}^{F};Z^T_b|W_b,\mathcal{C})\right)\nonumber\\
&\hspace{-0.6cm} = &\hspace{-0.5cm} \frac{1}{T}\Big(I(X_{\mathcal{S},b}^{T};Z^T_b|\mathcal{C})-H(R_{K,b}R_{K.b}^{F}|W_b,\mathcal{C})\nonumber\\
&&\hspace{3.7cm}+ H(R_{K,b}R_{K,b}^{F}|Z^T_b,W_b,\mathcal{C})\Big)\nonumber\\
&\hspace{-0.6cm}  \stackrel{(a)}{=} & \hspace{-0.5cm} \frac{1}{T}\Big(I(X_{\mathcal{S},b}^{T};Z^T_b|\mathcal{C}) -H(R_{K,b}R_{K,b}^{F})\nonumber\\
&&\hspace{3.7cm}+H(R_{K,b}R_{K,b}^{F}|Z^T_b,W_b,\mathcal{C})\Big)\nonumber\\
&\hspace{-0.6cm}  \stackrel{(b)}{\leq}&\hspace{-0.5cm}  \delta - \frac{1}{T}K\left(T \frac{\delta-R_f-\epsilon}{K}+T\frac{R_f}{K}\right)\nonumber\\
&&\hspace{3.7cm}+\frac{1}{T}H(R_{K,b}R_{K,b}^{F}|Z^T_b,W_b,\mathcal{C})\nonumber\\
&\hspace{-0.6cm}  \stackrel{(c)}{\leq}&\hspace{-0.5cm}   \epsilon_{T}^{\prime\prime}(b),\nonumber
\end{eqnarray}
\fi
where (a) is since both $R_{K,b}$ and $R_{K,b}^{F}$ are independent of $W_b$ and the codebook. (b) is since both keys are uniform, the first includes $K$ variables of $T(\frac{\delta-\epsilon}{K}-\frac{R_f}{K})$ bits each, and the second $K$ shares of $T\frac{R_f}{K}$ bits each. (c) follows from \cite[Section V.B]{cohen2016secure}. In short, given the true $K$ defectives, the codebook and her output $Z^T_b$, Eve sees a simple MAC channel, with the $K$ defectives as users, each having its bin as a set of codewords, and the channel is a simple Boolean MAC followed by a BEC. Due to the symmetric codebook construction, Eve operates at a rate slightly below her capacity. Hence, she can identify which codeword was selected for each defective item, hence identify both $R_{K,b}$ and $R_{K,b}^{F}$.

Similarly for $b=1$, when the mixer has no available key from the feedback, i.e., considering $R_f=0$ for the encoding of the first batch, it is easy to see from (b) above that,
\begin{equation*}
    \frac{1}{T}I(W_1;Z^{T}_{1}|\mathcal{C})\leq \epsilon_{T}^{\prime}.
\end{equation*}
To conclude, from both above bounds we have
\ifjsait
\begin{equation*}
 \frac{1}{TB} I(W_1,\ldots, W_B;Z^{T}_{1},\ldots,Z^{T}_{B}|\mathcal{C})
 \leq \frac{1}{B} \left(\epsilon_{T}^{\prime} + \sum_{b=2}^{B} \epsilon_{T}^{\prime\prime}(b)\right)= \frac{1}{B} \Big(\epsilon_{T}^{\prime} + \epsilon_{T}^{\prime\prime}(B-1)\Big),
\end{equation*}
\else
\begin{multline*}
 \frac{1}{TB} I(W_1,\ldots, W_B;Z^{T}_{1},\ldots,Z^{T}_{B}|\mathcal{C})\\
 \leq \frac{1}{B} \left(\epsilon_{T}^{\prime} + \sum_{b=2}^{B} \epsilon_{T}^{\prime\prime}(b)\right)= \frac{1}{B} \Big(\epsilon_{T}^{\prime} + \epsilon_{T}^{\prime\prime}(B-1)\Big),
\end{multline*}
\fi
where $\epsilon_{T}^{\prime} \to 0$ and $\frac{B-1}{B}\epsilon_{T}^{\prime\prime} \to 0$ as $T \to \infty$.                                  %
\section{Converse (Necessity)}\label{converse}
In this section, we derive a necessity bound on the required number of tests, assuming a rate-limited feedback from the legitimate user to the lab. Note that the converse result below is independent to any specific implementation, hence, $T$ below is the \emph{total number of tests} used, and the feedback is general, not necessarily random bits. 

First, using Fano's inequality, if $P_e\rightarrow 0$ when $T\rightarrow\infty$, then
\begin{equation}\label{eq:fano}
H(W|Y^T,F^T) \leq  \left[\log \binom{N}{K}\right]\epsilon_T,
\end{equation}
for some $\epsilon_T \rightarrow 0$ as $T\rightarrow\infty$. The secrecy constraint implies that
\begin{equation} \label{eq:R case 5 0}
I(W;Z^T)=T\gamma_{T},
\end{equation}
where $\gamma_{T}\rightarrow 0$ as $T\rightarrow \infty$. Consequently,
\begin{eqnarray}\label{eq:low1}
&& \hspace{-0.8cm} H(W)\nonumber\\
& = & H(W|Z^T)+I(W;Z^T)\nonumber\\
& \stackrel{(a)}{=} & H(W|Z^T)+T\gamma_{T}\nonumber\\
& = & I(W;Y^T,F^T|Z^T)+H(W|Y^T,Z^T,F^T)+ T\gamma_{T}\nonumber\\
& \stackrel{(b)}{\leq} & I(W;Y^T,F^T|Z^T)+\left[\log \binom{N}{K}\right]\epsilon_T+T\gamma_{T}\nonumber\\
& \stackrel{(c)}{\leq} & I(W;Y^T,F^T|Z^T)+\left[T-\log(1-P_e)\right]\epsilon_T+T\gamma_{T}\nonumber\\
& \stackrel{(d)}{=} & I(W;F^T|Z^T)+I(W;Y^T|F^T,Z^T)+T\upsilon_{T}\nonumber\\
& \leq  & H(F^T|Z^T)+I(W,\textbf{X}_{\mathcal{S}_{w}}^T;Y^T|F^T,Z^T)+T\upsilon_{T}, 
\end{eqnarray}
where (a) follows from \eqref{eq:R case 5 0}; (b) follows from Fano's inequality as given in \eqref{eq:fano}; (c) follows from \cite[Theorem 3.1]{baldassini2013capacity}, which bounds the number of tests required whether the algorithm is adaptive or not, and even without any security constraint; (d) is true for some $\upsilon_{T}$ such that $\upsilon_{T} \to 0$ as $T \to \infty$, since $\log(1-P_e) \to 0$ with $T$ as well.

The following recursive lemma is now required.
\begin{lemma}\label{rec. lemma}
For each $t\in \{1,\ldots T\}$,
\begin{multline*}
H(F^{t}|Z^{t})+I(W,\textbf{X}^{t}_{\mathcal{S}_{w}};Y^{t}|F^{t},Z^{t})
\\
 \leq  H(F^{t-1}|Z^{t-1})
 + I(W,\textbf{X}^{t-1}_{\mathcal{S}_{w}};Y^{t-1}|F^{t-1},Z^{t-1})
\\
+ H(F_{t}|W,\textbf{X}^{t-1}_{\mathcal{S}_{w}},F^{t-1},Z^{t-1})
+ I(\textbf{X}_{\mathcal{S}_{w},{t}};Y_{t}|Z_{t}).
\end{multline*}
\end{lemma}
This recursive lemma is a direct consequence of \cite[Lemma 1]{C8}. In \cite{C8}, it is considered for a communication problem, unlike the pool testing problem in this work. However, note that all the conditions given in \cite[Lemma 1]{C8} hold here, since the set of random variables involved is the same, and all connections (e.g., dependence) between the random variables are the same.
To continue, we use \Cref{rec. lemma} recursively starting from \eqref{eq:low1}:
\begin{eqnarray} \label{eq:R case 5 14}
&& \hspace{-0.8cm} H(W) \nonumber\\
& \leq & H(F^{T}|Z^{T})+I(W,\textbf{X}^{T}_{\mathcal{S}_{w}};Y^{T}|F^{T},Z^{T}) +T\upsilon_{T}\nonumber\\
& \leq & H(F^{T-1}|Z^{T-1})\nonumber\\
&& +I(W,\textbf{X}^{T-1}_{\mathcal{S}_{w}};Y^{T-1}|F^{T-1},Z^{T-1})\nonumber\\
&& +I(\textbf{X}_{\mathcal{S}_{w},{T}};Y_{T}|Z_{T})+H(F_{T})+T\upsilon_{T}\nonumber\\
& \leq & H(F^{T-2}|Z^{T-2})\nonumber\\
&& +I(W,\textbf{X}^{T-2}_{\mathcal{S}_{w}};Y^{T-2}|F^{T-2},Z^{T-2})\nonumber\\
&& +I(\textbf{X}_{\mathcal{S}_{w},{T-1}};Y_{T-1}|Z_{T-1})+H(F_{T-1})\nonumber\\
&& +I(\textbf{X}_{\mathcal{S}_{w},{T}};Y_{T}|Z_{T})+H(F_{T})+T\upsilon_{T}\nonumber\\
& \leq & \ldots\nonumber\\
& \leq & \sum_{i=1}^{T}I(X_{\mathcal{S}_{w},{i}};Y_{i}|Z_{i}) +\sum_{i=1}^{T}H(F_{i})+\upsilon_{T}.\nonumber
\end{eqnarray}
Thus,
\begin{equation*} \label{eq:R case 5 15}
H(W) \leq  \sum_{i=1}^{T}I(X_{\mathcal{S}_{w},{i}};Y_{i}|Z_{i}) +\sum_{i=1}^{T}H(F_{i})+T\upsilon_{T}.
\end{equation*}
We now use the constraint $\frac{1}{T}\sum_{i=1}^{T}\log(|F_{i}|)\leq R_{f}$ and normalize by $T$. We have,
\begin{equation*} \label{eq:R case 5 16}
\frac{1}{T}H(W) \leq  \frac{1}{T}\sum_{i=1}^{T}I(X_{\mathcal{S}_{w},{i}};Y_{i}|Z_{i})+R_{f}+\upsilon_{T}.
\end{equation*}
To conclude, the well-known technique of introducing a time sharing random variable is used. Assume $Q$ is independent of $\textbf{X}_{\mathcal{S}_{w}}^T,Y^{T},Z^{T}$ and uniform on $\{1,\ldots,T\}$, this results in
\begin{eqnarray*}
&& \hspace{-0.8cm} 
\frac{1}{T} H(W) \nonumber\\
&\leq& R_{f}+\frac{1}{T}\sum_{i=1}^{T}(I(X_{\mathcal{S}_{w},{i}};Y_{i}|Z_{i})+\upsilon_{T}\nonumber\\
&=& R_{f}+\frac{1}{T}\sum_{i=1}^{T}(I(X_{\mathcal{S}_{w},{i}};Y_{i}|Z_{i},Q=i) +\upsilon_{T}\nonumber\\
&=& R_{f}+I(X_{\mathcal{S}_{w},{Q}};Y_{Q}|Z_{Q},Q)+\upsilon_{T}\nonumber\\
&=& R_{f}+I(X_{\mathcal{S}_{w}};Y|Z,Q)+\upsilon_{T}\nonumber\\
&\leq& R_{f}+I(X_{\mathcal{S}_{w}},Q;Y|Z)+\upsilon_{T}\nonumber\\
&=& R_{f}+I(X_{\mathcal{S}_{w}};Y|Z)+\upsilon_{T},
\end{eqnarray*}
where $X_{\mathcal{S}_{w}}:=X_{\mathcal{S}_{w},{Q}}, Y:=Y_{Q}, Z:=Z_{Q}$.

Now, the upper bound we drive (in the asymptomatic regime) cannot be greater than the upper bound without the secrecy constraint \cite{shannon1956zero,aldridge2012adaptive}. Using the same techniques without the security constraint, 
\[
\frac{1}{T} H(W) \leq I(X_{\mathcal{S}_{w}};Y)+\epsilon_T. 
\]
From both above bounds,  
\begin{equation*}
\frac{1}{T} H(W) \leq  \min\{I(X_{\mathcal{S}_{w}};Y)+\epsilon_T, I(X_{\mathcal{S}_{w}};Y|Z)+R_{f}+\upsilon_{T}\}.
\end{equation*}
The eavesdropper can obtain information only via the outcomes of the legitimate user, i.e., $p(y,z|x) = p(y|x)p(z|y)$, thus we have
\ifjsait
\begin{equation*}
\frac{1}{T} H(W)\leq \min\{I(X_{\mathcal{S}_{w}};Y)+\epsilon_T,\\ I(X_{\mathcal{S}_w};Y)-I(X_{\mathcal{S}_w};Z)+R_{f}+\upsilon_{T}\}.
\end{equation*}
\else
\begin{multline*}
\frac{1}{T} H(W)\leq \min\{I(X_{\mathcal{S}_{w}};Y)+\epsilon_T,\\ I(X_{\mathcal{S}_w};Y)-I(X_{\mathcal{S}_w};Z)+R_{f}+\upsilon_{T}\}.
\end{multline*}
\fi
Since $H(W)=\log\binom{N}{K}$, we have
\ifjsait
\begin{eqnarray*}
  \frac{1}{T}\log\binom{N}{K} &\leq& \min\{I(X_{\mathcal{S}_{w}};Y)+\epsilon_T,I(X_{\mathcal{S}_w};Y) - I(X_{\mathcal{S}_w};Z)+ R_{f}+\upsilon_{T}\}\\
  & \leq & \min\{1+\epsilon_T, 1-\delta+R_f+\upsilon_T\}.
\end{eqnarray*}
\else 
\begin{eqnarray*}
  \frac{1}{T}\log\binom{N}{K} &\leq& \min\{I(X_{\mathcal{S}_{w}};Y)+\epsilon_T,\\
  && \hspace{0.6cm}I(X_{\mathcal{S}_w};Y) - I(X_{\mathcal{S}_w};Z)+ R_{f}+\upsilon_{T}\}\\
  & \leq & \min\{1+\epsilon_T, 1-\delta+R_f+\upsilon_T\}.
\end{eqnarray*}
\fi
That is,
\[
	T \ge  \frac{1}{\min\{1+\epsilon_T ,1-\delta+R_{f}+\upsilon_T}\}\log\binom{N}{K}.
\]
for some $\epsilon_T $ and $\upsilon_T$ such that $\epsilon_T \to 0$ and $\upsilon_T \to 0$ as $T \to \infty$.                                 %

\section{Secure Adaptive Group Testing with Curious but Non-malicious Non-cooperative Labs}\label{sec:labs}
In this section, we briefly consider two additional models of SAGT, in which we distinguish between the labs performing the tests and the legitimate doctor analyzing the tests results in order to identify the defective items. While the two models build on the techniques developed in the previous sections, they depict very practical scenarios, and, most importantly, show how data from one lab can be used to protect the others, resulting in a distributed testing procedure, where each lab performs a fraction of the tests yet is unable to gain any information on the items, an all this with the number of tests done in total equal to the one with no secrecy constraint.

In both models we assume curious yet non-malicious, non-cooperative labs which, in a sense, take the place of the eavesdropper. Specifically, the nurse Alice collects the blood samples of $N$ patients and prepares $T$ tubes using the testing matrix given in \Cref{LowerBound}. The nurse then divides the $T$ tubes into $1/ \delta$ sets of $T\delta$ pool-test each, and sends each set to a distinct lab (Eve). Each lab tests the $T\delta$ tubes received form the nurse and send the test results (namely which tube is positive) to the doctor's clinic. Thus, pool-tests are distributed to the labs to be checked, and we wish to keep the status of the items secret from these labs. This model is useful in practice, when the number of pool-tests collected by the nurse is large, thus we want to reduce the testing time by using multiple labs in parallel. 

In the first model, there is private feedback link from the doctor to the mixer. In the second model, this private feedback link is not available, yet there is a \emph{direct feedback} link from each each lab to the mixer. A graphical representation is given in \Cref{figure:group_testing_model_doctor}. 

The first model is in fact simple, given the results derived thus far. The doctor who gets the results from all the labs, uses the decoding algorithm given in \Cref{LowerBound}, and identifies the infected patients while, at the same time, sends symbols for the next batch of tests to the nurse over the private feedback link (green link in \Cref{figure:group_testing_model_doctor}). Reliability and secrecy are guaranteed using the test matrix proposed in \Cref{LowerBound}, since each lab has only $T\delta$ pool-tests.     

The second model \emph{cannot be viewed as a direct consequence of \Cref{LowerBound}, as the feedback is clearly visible to the lab sending it, hence it cannot be considered as a private feedback hidden from Eve}. Yet, each of the $1/\delta$ labs can send the results of the $T\delta$ tested tubes (rate permitting) directly to the nurse Alice over feedback link (orange link in \Cref{figure:group_testing_model_doctor}). Thus, the content on the link from each lab is obviously known to the lab which sent it. However, we still want to use the data obtained from these feedback links to confuses each lab regarding which patients are infected. 

The achievability with these ``locally-public" feedback\footref{public_link} links is based on the assumption that each lab knows only the information it sent to the nurse Alice. The remaining $T(1-\delta)$ outcomes can thus be used to \emph{create a random key to achieve security}. Particularly, the rate of the secret-sharing key which can be generated in each batch is
\ifjsait
\begin{equation*}
    R_{pf} \leq \min\Big\{ \left[I(X_{\mathcal{S}}^{T};Y^T) - I(X_{\mathcal{S}}^{T};Z^T)\right]^{+}, I(X_{\mathcal{S}}^{T};Z^T)\Big\}= \min\left\{ 1 - \delta, \delta \right\}.
\end{equation*}
\else
\begin{multline*}
    R_{pf} \leq \min\Big\{ \left[I(X_{\mathcal{S}}^{T};Y^T) - I(X_{\mathcal{S}}^{T};Z^T)\right]^{+},\\  I(X_{\mathcal{S}}^{T};Z^T)\Big\}= \min\left\{ 1 - \delta, \delta \right\}.
\end{multline*}
\fi
This is a direct consequence of the general proof given in \cite{C3} (see also \cite{gunduz2008secret}). However, unlike the solution proposed in \cite{C3,gunduz2008secret}, in the problem we consider in this section Alice and Bob need to extract a uniform random key (with independent bits) of length $TR_{pf}$ from the $T$ test outcomes. For this, before using the solution proposed in \Cref{LowerBound}, we will use Leftover Hash Lemma scheme \cite{haastad1993construction,barak2011leftover}. 

In the code construction phase, fix $T$ satisfying \Cref{direct_bab} and choose integers $F$ and $M$ such that
$$\log_2(F) = T(R_{pf}/K) \mbox{ and } \log_2(M) = T(\delta-R_{pf}-\epsilon)/K.$$
Then, generate a random codebook $\mathcal{C}_{1}$ in the same manner given in \Cref{LowerBound1}.

Now we describe the generation of the keys created with the information sent over the locally-public adaptive link from the labs to the nurse. Each lab uses this link to send the same information the lab sends to the doctor about the test results. Therefore, in a batch, both the nurse and the doctor receive $T$ bits from the $L$ labs. It is important to note that due to the random code construction, the specific input probability we use, and the fact that the set of defective items is uniformly distributed over all possible sets of size $K$, \emph{the test results are random, i.i.d., with a $\{1/2,1/2\}$ distribution}. We divide the key generation to two steps. In the first step, the nurse and the doctor need to extract uniformly random and independent key of length $TR_{pf}$ from the $T$ outcomes. Let $\mathbb{H}$ be a universal hash function. Hence, using the Leftover Hash Lemma \cite{haastad1993construction,barak2011leftover}, from the $T$ bits received in each batch, take $S = \mathbb{H}(Y^T)$, to obtain a (truncated) output of $TR_{pf}$ uniformly random and independent secret bits. In other words, from $1/\delta$ sets of $\delta T$ bits received from each lab, the mixer and the doctor can agree on $\delta T$ bits secret from all labs. Now, given $S$, the second step of the key generation is as in \Cref{LowerBound2}. Finally, the testing process, the reliability and the information leakage at the eavesdropper are a direct consequence of \Cref{LowerBound}. Hence, substituting $M=2^{T\frac{\delta-R_{pf}-\epsilon }{K}}$ in \Cref{direct lemma1} and replacing $\log\binom{N-K}{i}$ by it's asymptotic value $i\log N$, we have
\begin{equation*}
 T \geq  \frac{1+\varepsilon}{\min\{1,1-\delta+R_{pf}\}} K \log N,
\end{equation*}
for $R_{pf} \leq \min\left\{ 1 - \delta, \delta \right\}$.
\section{Secure Test Design Depending on the Outcome of Previous Pool-Test}\label{LowerBound_moel2}
In this solution, the outcome of each \emph{separate pool-test}, according to the rate of the private adaptive link, can be available at the mixer to influence the next test.
Since the feedback rate might be smaller than 1, we assume the legitimate parties can agree on which pool-tests depend on the symbols sent over the link. In other words, they can synchronize on when the feedback affects the mixing process. This assumption is not trivial, however, in various scenarios based on GT \cite{du2000combinatorial} this assumption holds, e.g., in the channel problem, when the encoder (mixer) and the decoder (lab) schedule the transmission over the main channel and the feedback.

\begin{figure}
  \centering
  \ifjsait
  \includegraphics[trim= 0cm 0cm 0cm 0cm,clip,scale=1.2]{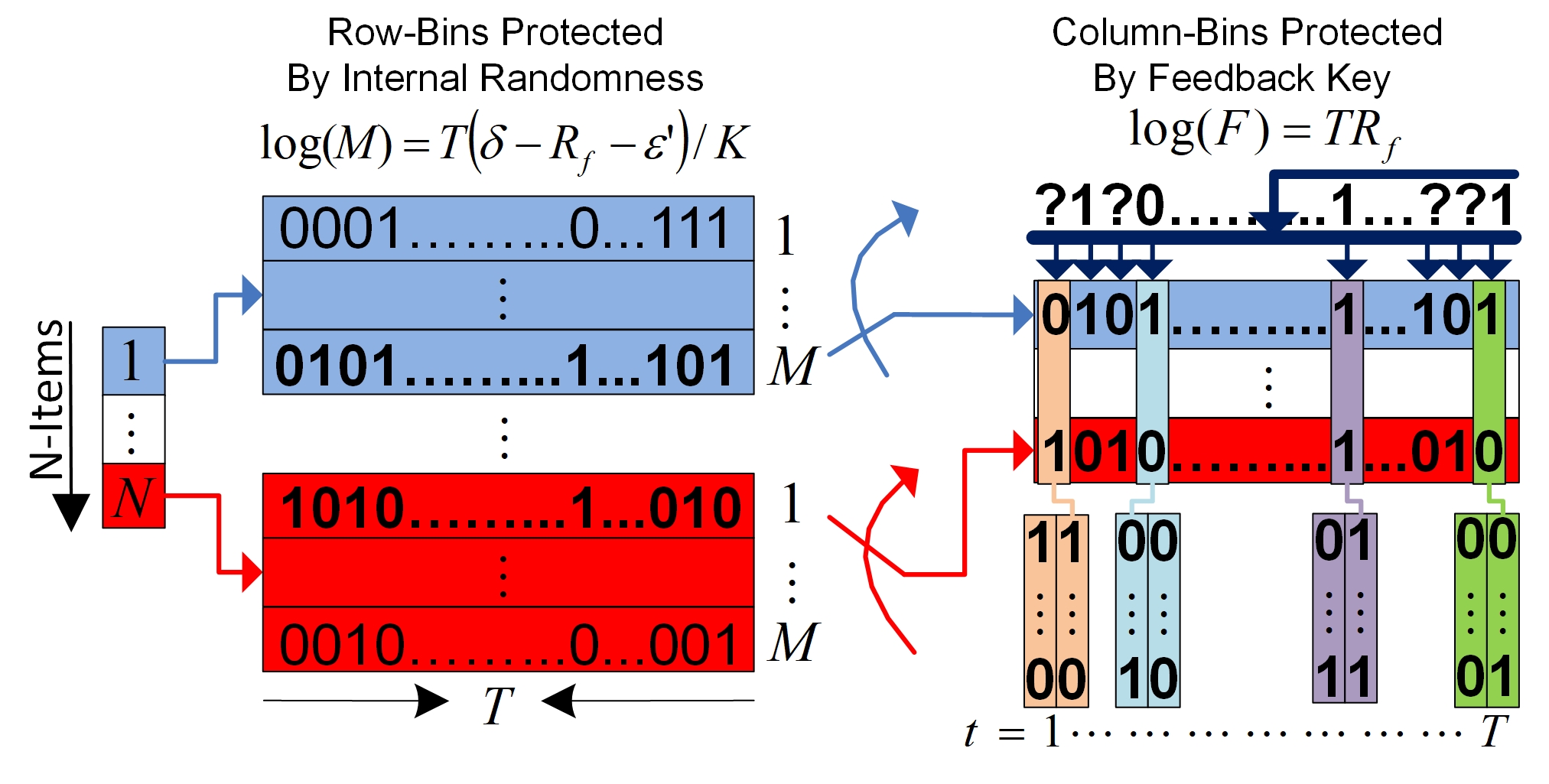}
  \else
  \includegraphics[trim= 0cm 0cm 0cm 0cm,clip,scale=1]{fig/Wiretap_coding5_one_col_AdpPerPoolS2.jpg}
  \fi
  \caption{Encoding process for a SAGT code where the test design depends on the outcome of previous pool-test.}
  \label{fig:WiretapCoding2}
  \vspace{-4mm}
\end{figure}

In the code construction phase, we generate the code as in \Cref{LowerBound}, yet, with one codeword in each sub-bin (i.e., the feedback will not be used to select the rows in the original testing matrix). Hence for each item we have a row-bin with $M$ row-codewords. Yet, for each possible column in the original testing matrix, we generate a column-bin with two column-codewords.
Then, for each $j$-th item, the mixer randomly select one row for the original testing matrix randomly as in the first setup, from the $j$-th row-bin, yet, per pool-test, one column-codeword is selected from its column-bin to finally define which items will participate in this pool-test. This column is selected according to the previous outcome of the feedback, sent after each separate pool-test by the legitimate lab to the mixer.
If the feedback of the previous outcome is not available to the mixer, since the rate of the feedback is limited and a bit is available only once per several tests, the mixer uses the column of the original testing matrix. For example, when the rate of the feedback, $R_f$, is 0.5, the mixer will use for each second pool-test, the information obtained from the feedback to select one column-codeword from the corresponding column-bin. In the remaining $0.5 \cdot T$ pool-test, the mixer will use the column of the original matrix.
\Cref{fig:WiretapCoding2} gives a graphical representation of the code.

Specifically, in the codebook generation phase, using a distribution $P(X^{T})=\prod^{T}_{i=1}P(x_i)$, for each item we generate a row-bin with $M$ independent and identically distributed row-codewords. Then in the same way, using a distribution $P(\tilde{X}^{N})=\prod^{N}_{i=1}P(\tilde{x}_i)$, for each possible column $x^{N}(t)$  we generate a column-bin with two independent and identically distributed column-codewords.

In the testing phase, for each item $j$, the mixer selects uniformly at random one row-codeword $x^{T}_{j}(m)$ from his row-bin.
The lab selects randomly one index in $\mathcal{I}\in \{1,\ldots,2^{TR_f}\}$, and before each pool-test $t\in T$, one bit of the index will be sent to the mixer over the feedback to influence the next test. Then, in each pool-test $t$, using the previous feedback outcome $F_{t}$, the mixer selects one column-codeword $\tilde{x}^{N}_{t}(F_{t})$ from the $x^{N}(t)$ column bin. In the case that the previous feedback outcome is not available at the mixer, the mixer uses the original column.

\ifveryshort
The decoder at the legitimate lab after each round know what was the indexes of the columns selected from the column-bins by the mixer for each pool-tests.
Hence, the decoder procedure, and the analysis of the reliability and the information leakage at the eavesdropper are almost a direct consequence of \Cref{LowerBound}.
\else
The decoder at the legitimate lab knows exactly which pool-test depends on the bits sent over the feedback and what were the indices of the columns selected from the column-bins by the mixer for each pool-test.
Hence, the decoding procedure after $T$ pool-tests, and the analysis of the reliability and the information leakage at the eavesdropper are a direct consequence of \Cref{LowerBound}.
\off{Hence, the decoder after $T$ pool-tests looks for a collection of $K$ codewords $X_{S_{\hat{w}}}^{T}$, \textit{one from each row-bin}, for which $Y^T$ is most likely. Namely, $P(Y^{T}|X_{S_{\hat{w}}}^{T})>P(Y^{T}|X_{S_{w}}^{T}), \forall w \neq \hat{w}$.
Then, the legitimate lab declares $\hat{W}(Y^T \times F)$ as the set of row-bins in which $\hat{w}$ reside.
The analysis of the reliability and the information leakage at the eavesdropper are almost a direct consequence of \Cref{LowerBound}.}
\fi
                             %
\section{Conclusions}\label{conc}
We proposed a new secure adaptive GT algorithm with a private adaptive feedback link of rate $R_f$. We show that in the GT setting, a novel secret sharing scheme is required for the information obtained over the feedback link. We proved that when there is an eavesdropper who may observe a fraction $\delta$ of the outcomes, the number of tests required for both correct reconstruction at the legitimate user, with high probability, and negligible mutual information at the eavesdropper is $1/min\{1,1-\delta+R_f\}$ times the number of tests required with no secrecy constraint. This was done by deriving both sufficiency and necessity bounds on the number of tests required to obtain both reliability and the secrecy constraint. Finally, we proposed two additional secure adaptive GT models using curious but non-malicious labs. One with a private adaptive feedback link and the second with a locally-public feedback link form each of the independent labs.                                %
\bibliographystyle{IEEEtran}
\bibliography{references}
\end{document}